\documentclass[letterpaper,twocolumn,10pt]{article}
\usepackage{usenix2019_v3}

\usepackage{times}
\usepackage{latexsym}
\usepackage[export]{adjustbox}
\usepackage{hyperref}
\usepackage{multirow}
\usepackage{tikz}
\usepackage{array}
\usepackage{xspace}
\usepackage{multirow}
\usepackage{comment}
\usepackage{url}
\usepackage{xurl}
\usepackage[normalem]{ulem}
\usepackage{mdwlist}
\usepackage{algorithm}
\usepackage{amssymb,amsmath}
\usepackage{booktabs}
\usepackage[noend]{algpseudocode}
\usepackage{centernot}
\usepackage{fancyvrb}
\usepackage{listings}
\usepackage{alltt}
\usepackage{subcaption}
\usepackage[ampersand]{easylist}
\usepackage{fancybox}
\usepackage{graphicx}
\usepackage{tcolorbox}
\usepackage{balance}
\usepackage{xcolor}
\usepackage{cite}

\definecolor{electricyellow}{rgb}{1.0, 1.0, 0.0}
\newcommand{\highlight}[1]{%
  \ooalign{\hss\makebox[0pt]{\fcolorbox{green!0}{electricyellow!40}{$#1$}}\hss\cr\phantom{$#1$}}%
}

\definecolor{light-gray}{gray}{0.8}
\definecolor{amethyst}{rgb}{0.6, 0.4, 0.8}

\makeatletter
\newcommand{\smallsym}[2]{#1{\mathpalette\make@small@sym{#2}}}
\newcommand{\make@small@sym}[2]{%
  \vcenter{\hbox{$\m@th\downgrade@style#1#2$}}%
}
\newcommand{\downgrade@style}[1]{%
  \ifx#1\displaystyle\scriptstyle\else
    \ifx#1\textstyle\scriptstyle\else
      \scriptscriptstyle
  \fi\fi
}

\newcommand{\HS}{\hspace{1mm}}

\newcommand\pagebudget[1]{}
{}
{}

{}

\newcommand{\sys}{\textsf{Poise}\xspace}

\hypersetup{
     bookmarks=true,
     colorlinks=true,
     linkcolor=black,
     citecolor=black,
     filecolor=black,
     urlcolor=black,
     pdfauthor = {},
     pdftitle = {},
     pdfkeywords = {}
}

\newenvironment{packeditemize}{
\begin{itemize}
  \setlength{\itemsep}{0.3pt}
  \setlength{\parskip}{2pt}
  \setlength{\parsep}{0pt}
}{\end{itemize}}

\global\let\shownotes\empty
\ifx \shownotes \empty
\newcommand\todo[1]{\textcolor{purple}{TODO: #1}}

\newcommand\missing[1]{\textcolor{red}{#1}}

\newcommand\removed[1]{\textcolor{red}{\sout{#1}}}

\else
\newcommand\todo[1]{}

\newcommand\missing[1]{}

\newcommand\removed[1]{}

\fi

\newcommand\OMIT[1]{}

\definecolor{orange}{rgb}{0.9,0.4,0.0}
\definecolor{purple}{rgb}{0.8,0,0.7}

\def\subsubsection{\@startsection{subsubsection}
	{3}
	{\z@}
	{0.1ex plus 0.1ex minus 0.1ex}
	{0ex}
	{\normalfont\normalsize\itshape\textbf}}
\makeatother

\begin{document}

\newcommand\blfootnote[1]{%
  \begingroup
  \renewcommand\thefootnote{}\footnote{#1}%
  \addtocounter{footnote}{-1}%
  \endgroup
}

\title{Programmable In-Network Security for Context-aware BYOD Policies \vspace{6mm}}
\author{{\rm Qiao Kang$^\dag$,} {\rm Lei Xue$^\star$}, {\rm Adam Morrison$^\dag$}, {\rm Yuxin Tang$^\dag$}, {\rm Ang Chen$^\dag$}, {\rm Xiapu Luo$^\star$}\\[0.2mm]
{\normalsize\rm $^\dag$Rice University, $^\star$The Hong Kong Polytechnic University}}

\maketitle
\sloppy
\thispagestyle{plain}
\pagestyle{plain}


\begin{abstract}
Bring Your Own Device (BYOD) has become the new norm in enterprise networks, but BYOD security remains a top concern.
Context-aware security, which enforces access control based on dynamic runtime context, holds much promise.
Recent work has developed SDN solutions to collect device context for network-wide access control in a central controller.
However, the central controller poses a bottleneck that can become an attack target, and processing context changes
at remote software has low agility.

We present a new paradigm, \textit{programmable in-network security} (\sys), which is enabled by the emergence
of programmable switches. At the heart of \sys is a novel switch primitive, which can be programmed to support
a wide range of context-aware policies in hardware. Users of \sys specify concise policies,
and \sys compiles them into different instantiations of the security primitive in P4.
Compared to centralized SDN defenses, \sys is resilient to control plane saturation attacks, and it dramatically
increases defense agility.

\end{abstract}

\vspace{-1mm}
\section{Introduction}
\vspace{-1mm}
\label{sec:intro}

BYOD refers to the practice where enterprise employees could use privately owned
tablets, phones, and laptops at work~\cite{byod}. 
This practice has become the new norm in many organizations~\cite{byod-ibm,byod-companies,samsung-byod,ibm-byod-support,cisco-byod,byod-market-reports}, 
and its market is projected to grow and exceed \$73 billion by 2021~\cite{byod-market-reports}.

One of the top concerns, however, is BYOD security. As BYOD devices are generally less well managed
than their enterprise counterparts, they are easier targets to compromise~\cite{bring-your-own-risk,byod-threats,Dang-Pham-2015-comsec,Zahadat-2015-comsec}. 
This is further exacerbated by the fact that such devices are used to access sensitive
enterprise resources as well as untrustworthy services in the wild~\cite{benefits-and-risks,rise-and-risk}.
At its core, BYOD security represents a concrete instance of a fundamental challenge, sometimes known as
the ``end node problem''~\cite{end-node-problem-1,end-node-problem-2}. The ``end nodes'' are not subject to the
same level of centralized control, management, and protection as the enterprise infrastructure---we can 
easily update the access control lists on the gateway router, or patch newly discovered vulnerabilities
on a server, but ensuring that scattered end points are properly patched is much more difficult.
As such, insecure end devices tend to become the weakest link in the security chain~\cite{byod-weakestlink}.

One promising approach to BYOD security is to use \textit{context-aware} policies, which enforce access control based on devices' runtime context~\cite{hong-2016-pbs}. 
For instance, a policy may deny access from devices whose TLS libraries have not been updated~\cite{yu-2017-psi},
or grant access to devices that are physically located in the enterprise boundary~\cite{nils-2016-lbac},
or only allow the use of a sensitive service only if administrators are online~\cite{georgiadis-2001-teambased,tolone-2005-collaborative}. 
In each of these scenarios, we desire to make security decisions based on additional ``threat signals'', such as 
the device location, library version, or even the status of other devices in the network.
Context-aware policies are in a class of their own---they are much more \textit{dynamic}, as context can change frequently (e.g., GPS location),
and they require \textit{global visibility} of the entire network (e.g., administrators online).

Supporting context-aware policies in enterprise networks presents interesting research challenges. 
Some traditional systems operate at the server side~\cite{vmware-nsx,ward-2014-beyondcorp}, 
which enables easier management and update of security policies; others operate at the client side~\cite{wang-2015-deepdroid},
making it easier to access device context as threat signals. 
A common limitation, however, is that the individual nodes---clients or servers---only have local visibility, and such a ``tunnel vision''
hinders the ability to make synchronized security decisions network-wide~\cite{synchronized-security}. 
Latest proposals address this using centralized SDN, where a software controller collects context signals from all devices and enforces 
network-wide access control policies~\cite{hong-2016-pbs}. However, the central controller is vulnerable to control plane 
saturation attacks~\cite{shin-2013-avantguard}, and responding to threat signals in remote software incurs delay and decreases agility. 

\textbf{Our contribution.} 
We present a novel design called \sys, or \underline{p}r\underline{o}grammable \underline{i}n-network \underline{se}curity, 
whose goal is to address the limitations of centralized SDN defense. Instead of collecting and processing context signals at a remote 
software controller, we design a new security primitive that runs in switch hardware. This primitive can extract context signals 
from client traffic and change defense postures at hardware speeds, while maintaining full linespeed (100Gbps per switch port) packet processing. 
Moreover, it can be re-programmed by a declarative language to support a wide range of context-aware policies. 
This is achieved by interposing a policy compiler, which can compile policies to different instantiations of the security primitive. 
Compared to centralized SDN defenses~\cite{hong-2016-pbs,oconnor-2018-pivotwall,shin-2013-avantguard}, 
\sys enforces context-aware security without software processing at remote controllers, resulting in 
one-of-a-kind defenses that are efficient, programmable, highly agile, and resilient to control plane saturation attacks~\cite{shin-2013-avantguard}.

The key enabler for \sys is the emerging  \textit{programmable data planes} in latest networking technology. 
New switches, such as Intel FlexPipe~\cite{flexpipe}, Cavium XPliant~\cite{cavium}, and Barefoot Tofino~\cite{tofino}, 
can be programmed in P4~\cite{p4-lang} to support user-defined network protocols, custom header processing, and sophisticated state in hardware. 
P4 networks represent a major step beyond OpenFlow-based SDN: OpenFlow switches have fixed-function hardware, and they 
can only support programmable forwarding by parsimoniously invoking remote software controllers; but P4 switches offer 
unprecedented programmability that we can apply to every single packet without performance compromise. 
The novelty of \sys lies in leveraging the new hardware features for context-aware security---we encode context signals 
with user-defined protocols, make access control decisions using programmable packet processing, and support network-wide 
security policies by designing hardware data structures. 

\vspace{1mm}
After describing more background in \S\ref{sec:background}, we present: 
\vspace{-0.5mm}
\begin{packeditemize}
    \item The concept of programmable in-network security (\S\ref{sec:overview});
    \item A language and compiler for context-aware policies (\S\ref{sec:lang});
    \item A novel in-network security primitive, \sys (\S\ref{sec:primitive});
    \item The \sys orchestration service and device module (\S\ref{sec:orchestration});
    \item Prototype and evaluation of \sys that demonstrate its practicality, as well as its higher resilience to control plane saturation attacks
          and increased defense agility compared with OpenFlow-based defense (\S\ref{sec:eval});
\end{packeditemize}
\vspace{-0.5mm}
We then describe related work in \S\ref{sec:related}, and conclude in \S\ref{sec:conclusion}.

\vspace{-2mm}
\section{Background}
\label{sec:background}
\vspace{-1mm}

Context-aware security (CAS) stands in stark contrast to conventional security mechanisms---existing mechanisms
can only support  \textit{static} policies, but CAS uses \textit{dynamic} policies based on runtime context.
For instance, NAC (Network Access Control) mechanisms such as IEEE 802.1x~\cite{ieee-802.1x} and Cisco ACL~\cite{acl-cisco}
statically configure access control policies for a device or an IP prefix, respectively. Role- or attribute-based
access control mechanisms~\cite{ferraiolo-2007-rbac,parducci-2005-abac,ferraiolo-2001-rbac} also regulate access based
on statically defined roles or attributes. CAS, on the other hand, uses the broader context of a request as threat signals
(e.g., location/time of access, the status of the network), and whenever the signals change, the security posture would adapt accordingly.
The theoretical underpinnings of CAS have been studied more than a decade back~\cite{barth-2006-contextualintegrity},
and it recently found an array of new applications in securing IoT and mobile devices~\cite{fernandes-2016-flowfence,hong-2016-pbs,wang-2015-deepdroid,
jia-2016-contextiot,apthorpe-2018-contextualintegrity}. These devices, just like the BYOD clients in our scenario,
suffer from the ``end node problem''~\cite{end-node-problem-1,end-node-problem-2}.
CAS has proven to be effective for such scenarios, because it can enable a more precise protection based on threat signals
collected from the end nodes.

\vspace{-2mm}
\subsection{Design space}
\label{subsec:designspace}
\vspace{-1mm}

The concept of CAS by itself does not necessitate a  client-, server-, or network-based design; rather, these design points have pros and cons.
First off, purely server-side solutions are often ineffective, as we desire to collect context signals from client devices at runtime.
Therefore, typical CAS systems~\cite{hong-2016-pbs,wang-2015-deepdroid} need to install a \textit{context collection} module at the clients to
collect context signals. In terms of \textit{policy enforcement}, one could co-locate enforcement with context collection, resulting in a
purely client-based solution~\cite{wang-2015-deepdroid}. The main drawbacks, however, are that a) individual devices only have local views,
making network-wide decisions hard to come by, and that b) policy management is much harder, as policies are distributed to each device;
this might raise additional concerns if some policies are themselves sensitive data.

The latest response from the security community is to use network-based designs using a centralized SDN controller~\cite{hong-2016-pbs}.
The policy enforcement module runs as an ``SDN app'' in a remote controller, which collects context signals from all devices
and enforces access control in a centralized manner. This enables a network-wide view for holistic protection, and enables centralized policy management.
The design of \sys adopts similar security principles, although it has a key difference---the policy enforcement module runs in
data plane hardware, rather than control plane software.

\vspace{-2mm}
\subsection{Trust model}
\vspace{-1mm}

\sys adopts the same trust model from existing CAS solutions~\cite{hong-2016-pbs,wang-2015-deepdroid}.
Although existing solutions (and \sys) primarily focus on Android devices, the conceptual designs are generally applicable.

\noindent\textbf{CAS modules.}
The context collection and policy enforcement modules are trusted. The former runs in client devices, and the latter may run at
the client side (e.g., DeepDroid~\cite{wang-2015-deepdroid}), as an SDN app (e.g., PBS~\cite{hong-2016-pbs}), or inside the switch (\sys).
The context collection module can be installed as a privileged module on Android devices with OEM support,
which is a common practice in Enterprise Mobility Management (EMM) solutions~\cite{wang-2015-deepdroid,blackberry-emm,
vmware-airwatch-emm,symantec-emm}. This module is only activated when devices connect to the enterprise network with user consent,
which is supported by standard BYOD frameworks such as the work profile in Android for Work~\cite{android-for-work} or containers in Knox~\cite{samsung-knox}.
Users can install their favorite apps; some of these may be malicious and are the goal of CAS protection.
We also adopt the assumptions from existing work~\cite{wang-2015-deepdroid,hong-2016-pbs} that malicious apps cannot compromise the kernel
or obtain root privileges, i.e., the Android kernel and firmware are trusted.
While it is possible to relax these assumptions using trusted hardware~\cite{verified-boot,nauman-2010-remoteattestation,trustzone}
or privilege separation techniques~\cite{RootGuard-2014,dautenhahn-2015-nestedkernel}, these are an orthogonal line of work.

\noindent\textbf{Authentication.}
Resource \textit{authorization} mechanisms like \sys and PBS~\cite{hong-2016-pbs} are also orthogonal to the choice of \textit{authentication} techniques.
Common practices in enterprise networks, such as the SAE (simultaneous authentication of equals) protocol~\cite{harkins-2008-sae} in WPA3~\cite{wpa3},
or two-factor authentication using TOTP~\cite{m2011rfc}, are compatible. Only authenticated devices can further access enterprise resources under CAS.

\noindent\textbf{Context integrity and privacy.}
Network-based designs, such as PBS~\cite{hong-2016-pbs} and \sys, require context signals to be sent to the network.
The integrity and privacy of context packets are subjected to the same level of protection as normal traffic.
Communication between client devices and access points could be protected by WPA3~\cite{wpa3}, which encrypts traffic and
prevents masquerading or replay attacks. Communication between access points and the enterprise network, as well as within the network, could be protected by
MACsec~\cite{macsec}, which encrypts traffic at the Ethernet layer and is supported by commodity off-the-shelf switches,
such as Aruba 200~\cite{aruba220} and Cisco Nexus 3400~\cite{nexus3400} (the latter is P4-programmable).
\sys switches drop context packets immediately after processing them, so they will not be propagated further.

\vspace{-2mm}
\section{Programmable In-Network Security}
\label{sec:overview}
\vspace{-2mm}

This section describes why traditional network cannot support CAS, how OpenFlow-based SDN 
provides a partial solution, and how \sys can leverage programmable data planes to achieve a new design. 

\vspace{-2mm}
\subsection{Traditional networks are not enough}
\label{subsec:traditional}
\vspace{-1mm}

One natural place for implementing network access control is on network devices, e.g., switches or middleboxes. 
However, this is infeasible because traditional devices are customized for specific purposes, and they cannot 
be re-programmed to support new functions, such as a diverse set of CAS policies. 
For instance, switches speak TCP/IP, but they cannot understand
context information, such as GPS location, time of access, or library versions. We could in principle deploy specialized
middleboxes, but such devices are blackbox in nature and unamenable to change---policy updates
would be constrained by the speed of hardware upgrades, which is much slower than the speed needed for changing security postures.
As a result, traditional in-network security mechanisms merely provide \textit{fixed-function security}, such as 
static access control lists, firewalls, traffic filters, and and deep packet inspection (DPI). 
There is a fundamental gap between the dynamic nature of CAS and the static network devices. 

\vspace{-2mm}
\subsection{How about OpenFlow-based SDN?}
\label{subsec:openflow}
\vspace{-1mm}

Software-defined networking (SDN), a concept first proposed a decade ago~\cite{mckeown-2008-openflow}, has now become a reality.
Commercial deployments of SDNs are already in large scale---exemplary networks
include Microsoft's SWAN~\cite{hong-2013-swan} and Google's B4~\cite{jain-2014-b4}.
SDN networks offer \textit{control plane programmability} by the use of a software controller, which can be implemented in
general-purpose programming languages (e.g., OpenDaylight\cite{OpenDaylight}/Beacon\cite{Beacon} in Java, POX\cite{POX} in Python, and NOX\cite{NOX} in C++). 
Although the OpenFlow switch hardware remains fixed in function, they can send \texttt{PacketIn} messages to the central controller for 
programmable decisions. 
Control plane programmability underlies many recent developments in network security~\cite{oconnor-2018-pivotwall,shin-2013-avantguard,
shin-2012-cloudwatcher,hong-2016-pbs,shin-2013-fresco,porass-2012-fortnox}. 
In particular, PBS~\cite{hong-2016-pbs} uses a centralized SDN controller for context-aware security. 

However, in traditional SDN, programmability comes at a great cost, as it resides in a centralized software controller. 
First, \texttt{PacketIn} messages incur a round-trip delay between the switch and the remote controller,
whereas packets on the data plane fast path are directly processed at hardware speeds.
As such, we can only programmatically process a small set of packets---typically one packet per flow (e.g., the first packet).
Second, traditional SDNs are vulnerable to control plane saturation attacks~\cite{shin-2013-avantguard},
where an adversary can cause high-volume traffic to be sent to the software controller. 
A recent work OFX~\cite{sonchack-2016-ofx} has further highlighted that, for security applications that require dynamic, 
fine-grained decisions, centralized SDN defenses would pose a severe bottleneck. 
The key goal of \sys is to address the limitations of centralized SDN defense by enforcing CAS in switch hardware. 

\vspace{-2mm}
\subsection{Opportunity: Programmable data planes}
\vspace{-1mm}

\begin{figure}[t!]
\centering\includegraphics[width=7.2cm]{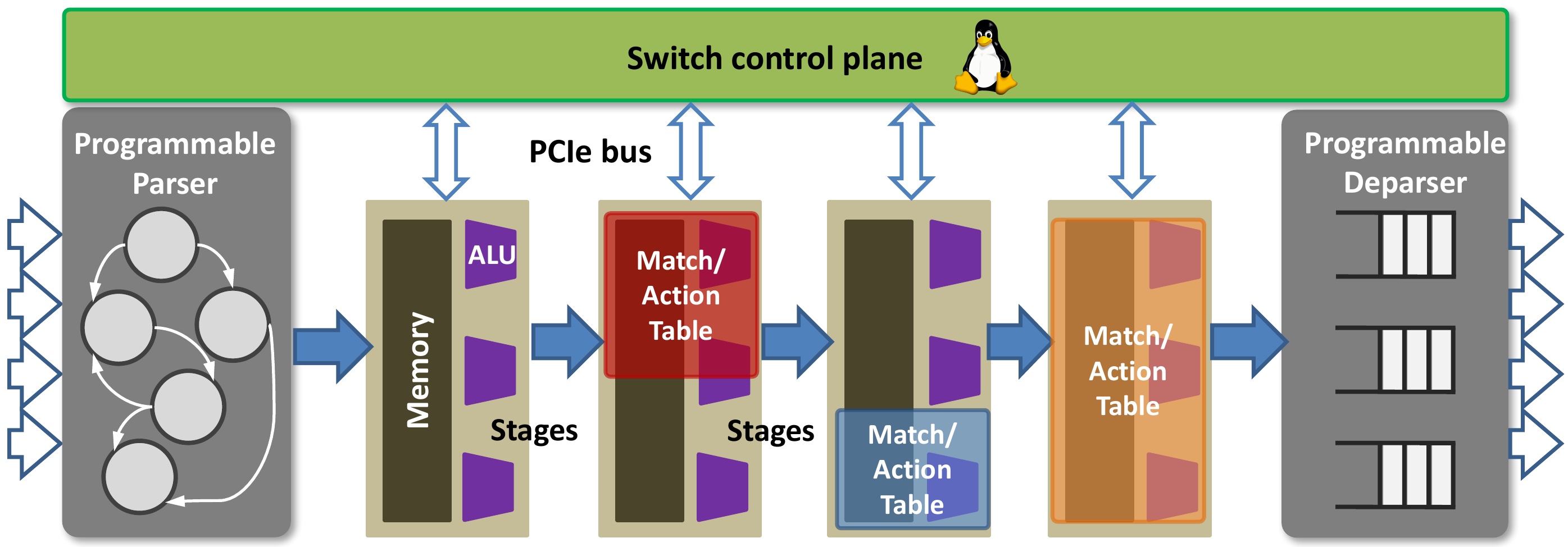}
\vspace{-1ex}
\caption{P4 switches are programmable in hardware. Packets first go through a programmable \textit{parser}, 
which supports user-defined protocols. Packet headers are then streamed through a number of hardware \textit{stages}, 
each of which contains \textit{registers} in memory, arithmetic logic units (\textit{ALUs}), and \textit{match/action tables}. 
Packets can be \textit{recirculated} to go through the stages multiple times to trigger different programmable elements.}
\vspace{-3ex}
\label{fig:pdp}
\end{figure}

\textit{Data plane programmability} represents the latest step in the networking community. In contrast to OpenFlow-based SDN, 
P4 networks are programmable in hardware. Figure~\ref{fig:pdp} illustrates the new features of programmable data planes. 
The key novelty of \sys is to leverage these features for security. 

\textbf{1. Customized header support for CAS.}
First, they can recognize customized protocols and headers beyond TCP/IP via the use of a programmable parser, without the need for hardware upgrades. 
Our observation is that this allows us to programmatically define context signals as special header fields, and embed them to network traffic. 
P4 switches can directly parse context signals from client traffic. 

\textbf{2. Linespeed processing for fastpath security decisions.}
Second, each hardware stage is integrated with ALUs (Arithmetic Logic Units) that can perform computation over header fields at linespeed.
The implication for security is that, without involving a remote software controller,
switches can evaluate context headers (e.g., GPS locations) and make security decisions
(e.g., location-based access control) on the fast path.

\textbf{3. Cross-packet state for network-wide security.}
Last but not least, the hardware stages also have persistent memory in read/write registers, and they can process packets based on persistent state. 
We observe that this enables the network to make coordinated
security decisions in a network-wide manner---decisions for one client could depend on past network behaviors, or activities from
other parts of the network. 

\vspace{1mm} These hardware features are programmable in P4~\cite{p4-lang,bosshart-2014-p4}. 
Switch programs can be compiled and installed from the switch control plane (Figure~\ref{fig:pdp}), which typically runs a customized 
version of Linux and has general-purpose CPUs. The P4 compiler maps a switch program to the available hardware resources~\cite{jose-2015-compiling}: 
programs that successfully compile on a target are guaranteed to run at linespeed, due to the pipelined nature of the hardware; 
programs that exceed available hardware resources would be rejected by the P4 compiler.

\vspace{-2mm}
\subsection{\sys overview}
\vspace{-1mm}

\begin{figure}[t!]
\centering\includegraphics[width=7.2cm]{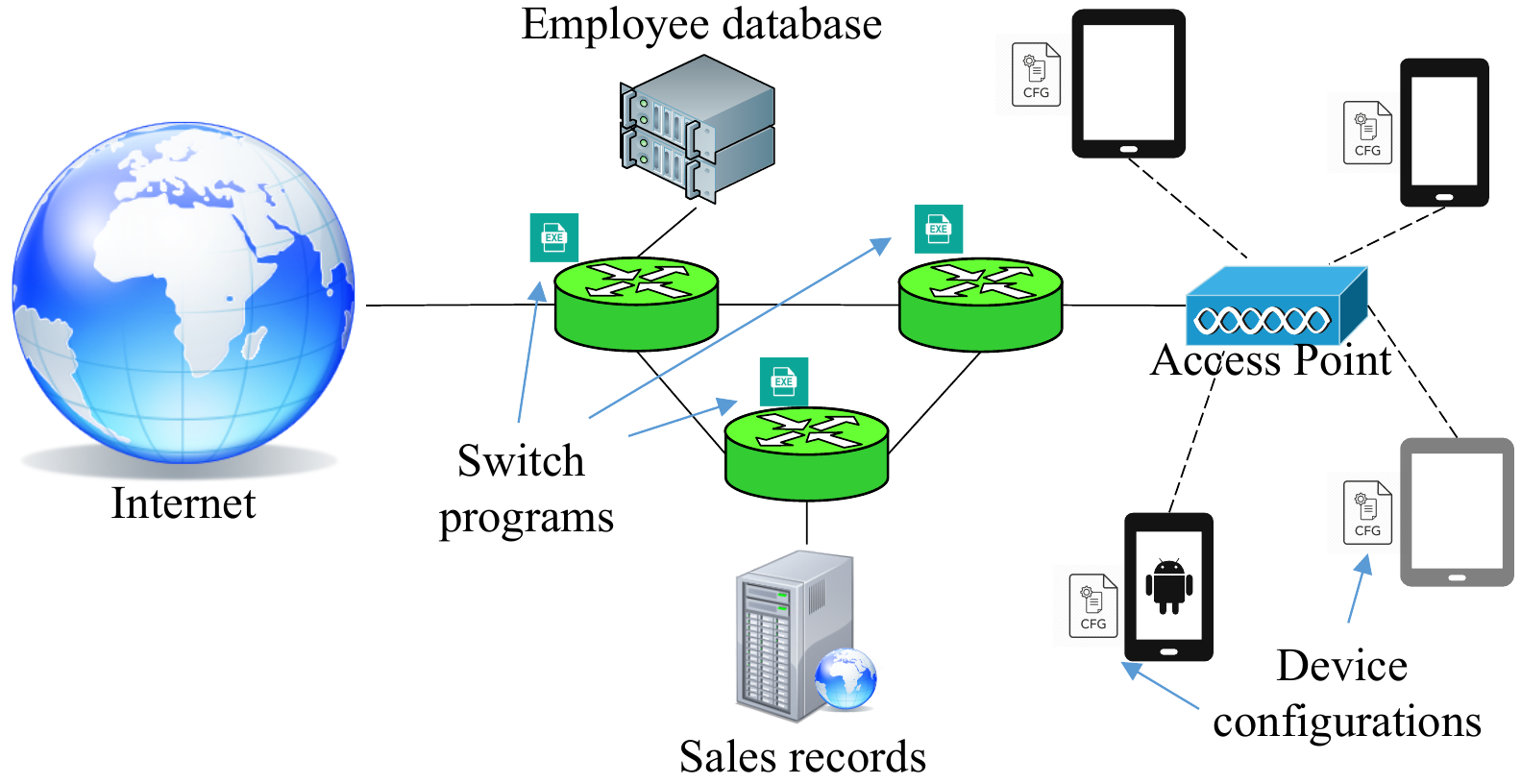}
\vspace{-1ex}
\caption{\sys compiles a high-level policy into a) switch programs, and b) device configurations, and enforces the policy inside the network.}
\vspace{-3ex}
\label{fig:scenario}
\end{figure}

Consider the enterprise network shown in Figure~\ref{fig:scenario}, which hosts several types of private data, such as employee
records and sales records, and also provides connectivity to the Internet. The operator wants to enforce dynamic access control of sensitive
enterprise data in the presence of BYOD clients.
Instead of traditional, password-based access control, the policy might additionally specify that a) sales records should only be accessed
by devices belonging to the sales department; b) during regular work hours; c) from devices that are properly patched to address some
recently discovered vulnerability; and, d) a device can only access the sales records if the sales manager is online.
\sys is designed for advanced enterprise security policies such as these.

At the heart of \sys is a novel switch primitive that achieves context-aware, linespeed, in-network security on programmable data planes.
The design of this primitive also tackles a practical challenge in P4 programming. 
Since P4 programs specify low-level packet processing behaviors, they are akin to ``microcode'' programming and fairly challenging to write. 
Moreover, one often needs to hand-optimize P4 programs, e.g., merging multiple match/action tables, to reduce resource usage. 
Therefore, we allow network operators to specify context-aware security policies in a declarative language that is 
much higher-level than P4. Our \sys compiler can then generate optimized P4 programs to enforce the policies. 
These programs are different instantiations of the security primitive, which can make policy-specific decisions. 
The \sys compiler also generates configurations for the context collection module, which runs in Android clients as a pre-positioned kernel module. 
It collects context signals based on the configuration, and sends out periodic context packets to the network. 
Policy changes can be easily supported by a reconfiguration. Client configurations may not be affected by policy updates, 
unless the new policies require new types of context signals to be collected. 

Next, we first describe how \sys presents a declarative policy interface, and how \sys compiles the key policy constructs in \S\ref{sec:lang}. 
We then describe the programmable \sys primitive in \S\ref{sec:primitive}, and the client module in \S\ref{sec:orchestration}.

\vspace{-2mm}
\section{The \sys Language and Compiler}
\label{sec:lang}
\vspace{-2mm}

The policy language in \sys is inspired by the Frenetic family of SDN programming languages~\cite{foster-2011-frenetic,monsanto-2012-netcore,
anderson-2014-netkat,monsanto-2013-pyretic,schelsinger-2014-concurrentnetcore}, but we adapt them a) from an OpenFlow setting to P4, which
supports richer header operations and state, and b) from a network management setting to security, by supporting security context.
Specifically, we have designed the \sys language based on Pyretic NetCore~\cite{monsanto-2013-pyretic}, which models network policies
as a series of match/action statements. In terms of the semantics of the language, a policy represents a function that maps an incoming
packet to zero (i.e., drop), one (i.e., unicast), or more (i.e., multicast) outgoing packets\footnote{We refer interested readers
to \cite{foster-2011-frenetic} for a more formal treatment of the Frenetic language semantics.}. A policy could be as simple as \texttt{drop},
which drops all packets, although practically, the policy would make a decision based on the context a packet carries, such as
\texttt{if match(dip==66.220.144.0) then drop}, which blacklists a block of destination IP addresses, or \texttt{if match(0800<=time<=1800) then drop},
which denies access depending on the time of day.
Figure~\ref{fig:grammar} summarizes the language syntax, and the highlighted portions show the differences
from NetCore, which we explain more below.

\vspace{-2mm}
\subsection{Key language constructs}
\vspace{-1mm}

\noindent\textbf{Security context.} \sys supports CAS by allowing custom definitions of security context,
such as the \texttt{time} field mentioned above, and the \texttt{dev} field mentioned later. \sys treats such context
as special header fields with a default 32-bit width, although the field sizes can be further specified by the policy programmer for customization.
When a policy refers to multiple types of context, \sys structures the context headers in the order in which they appear in the policy program.

\vspace{1mm}\noindent\textbf{Context operations.} \sys also supports sophisticated operations over context headers, as indicated in the
expressions and predicates in Figure~\ref{fig:grammar}. An expression could be a constant, an arithmetic operation over header fields,
or a complex expression over subexpressions. Security decisions are made based on predicates over expressions,
where the $\circ$ operator indicates comparisons such as $>$, $<$, and so on.
Context can also be tested against global, constant lists, which are pre-defined in the policy to encode membership relations.
For instance, one could define a list of devices with administrative roles as \texttt{def adminlst = ["dev1", "dev2"]}.
Then, the policy could refer to the lists as part of the decision-making process, such as
\texttt{if match(!dev in adminlst) then fwd(mbox)}, which forwards traffic from non-admin devices to a middlebox for traffic scrubbing.
We note that the original NetCore does not support the use of context or sophisticated context operations;
rather, \sys adds such extensions based on the extra processing power in P4 for security support.

\begin{figure}[t!]
  \centering
  \small
  \[ \begin{array}{rclr}
    \multicolumn{3}{l}{\textbf{Primitive Actions}} \\
    A &::=   & \mathsf{drop \HS | \HS fwd(port) \HS | \HS flood \HS | \HS log} \\
    \multicolumn{3}{l}{\textbf{Expressions}} \\
    E & ::=   & \highlight{\mathsf{v \HS | \HS e_1+e_2 \HS | \HS e_1-e_2 | \HS e_1*e_2}} \HS | \HS M\\
    \multicolumn{3}{l}{\textbf{Constant Lists}} \\
    L &::=   & \highlight{\mathsf{[v]}} \\
    \multicolumn{3}{l}{\textbf{Predicates}} \\
      &      & \highlight{\mathsf{match(h\HS in \HS l)} } \HS | \HS \mathsf{P\&P \HS | \HS (P|P) \HS | \HS !P} \\
    \multicolumn{3}{l}{\textbf{Monitors}} \\
    M &::=   & \mathsf{count(P)} \\
    \multicolumn{3}{l}{\textbf{Policies}} \\
    C &::=&  \mathsf{A \HS | \HS if \HS P \HS then \HS C \HS else \HS C \HS | \HS (C | C) \HS} \\
  \end{array} \]
  \vspace{-4ex}
  \caption{The language syntax for \sys policies. Context fields are represented as $\mathsf{h}$.
           Expressions are represented as $\mathsf{e}$, or $\mathsf{v}$ (constants).
           The $\circ$ operator indicates comparisons.}
  \vspace{-4ex}
  \label{fig:grammar}
\end{figure}

\vspace{1mm}\noindent\textbf{Stateful monitors.} Unlike NetCore, \sys supports stateful policies which make security decisions based on
network-wide state. This is done via a monitor expression, which monitors activities of interest and maintains state for decision making.
A monitor expression is written as \texttt{count(pred)}, which counts the number of packets that satisfy the predicate \texttt{pred} in the
current time window; for instance, \texttt{count(match(is\_admin))} counts the number of packets generated from a device with an administrative
role. The counters are periodically reset to zero when a new time window begins. These monitors enable programmers to write
network-wide policies, where the processing of a packet depends on not only its own context, but also the context of other traffic.
This is different from stateless NetCore policies, where monitors passively collect traffic statistics, but
do not affect how forwarding decisions will be made.

\vspace{1mm}\noindent\textbf{Actions.} The final decision of a \sys policy is represented by its action field. Currently,
\sys supports four types of actions. The \texttt{drop} decision denies access. The \texttt{fwd} decision allows access, and
can be further parameterized by an outgoing switch port, so that it can actuate further processing---e.g.,
sending packets through an DPI device that can be reached via a particular port. The \texttt{flood} decision broadcasts a
packet. The \texttt{log} decision sends a packet to a logger that detects potentially suspicious activity; this is achieved by aliasing the \texttt{fwd}
decision and specifying a special port for the local switch CPU. Packets sent for logging will be pumped to the control plane
of the switch, which maintains a logger. This can be easily generalized to enable remote logging, e.g.,
by wrapping the packet inside another IP header, where the destination IP represents a network activity logger.

\vspace{1mm}\noindent\textbf{Composing policies.}
Similar as NetCore, \sys can compose multiple policies \texttt{P1|P2|...|Pn} and compile them into a single switch program.
This is useful, e.g., when \texttt{Pi} and \texttt{Pj} check different context signals and the enterprise wants to apply them
in combination. The \sys compiler rejects the composition of conflicting policies at compilation time.

\vspace{-2mm}
\subsection{Example policies}
\label{sec:policy-example}
\vspace{-1mm}

The \sys language is expressive enough to capture a wide range of existing and new policies, and it is much more concise than low-level
languages such as P4. Next, we describe seven practical BYOD policies, where the first two are adapted from existing work~\cite{hong-2016-pbs} and
the rest are new policies supported by \sys. Variables \texttt{dev}, \texttt{time}, \texttt{lat}, \texttt{lon}, and \texttt{usr}
are customized header fields.

\newsavebox{\mybox}

\textit{P1: Block certain services in work hours~\cite{hong-2016-pbs}:}
A common BYOD policy is to block access from certain devices to entertainment websites during work hours:

\begin{lrbox}{\mybox}%
\begin{lstlisting}
  def businesslst = ["dev1", "dev2"]
  if match(dip==66.220.144.0 &
     dev in businesslst &
     (time>=0800)&(time<=1800))
  then drop
\end{lstlisting}
\end{lrbox}
\scalebox{0.8}{\usebox{\mybox}}
\vspace{-3mm}

\textit{P2: Direct traffic from guest devices through a middlebox~\cite{hong-2016-pbs}:} Another useful policy is
to distinguish traffic from authorized devices and guest devices, and direct guest traffic through a middlebox
for traffic scrubbing:

\begin{lrbox}{\mybox}%
\begin{lstlisting}
  def authlst = ["dev1", "dev2"]
  if match(dev in authlst)
  then  fwd(server)
  else  fwd(mbox)
\end{lstlisting}
\end{lrbox}
\scalebox{0.8}{\usebox{\mybox}}
\vspace{-3mm}

\noindent\textbf{New policies.}
There are also useful policies in \sys that cannot be easily supported in traditional networks; they are implementable in \sys due to the use
of programmable data planes, which can perform arithmetic operations over context headers, and maintain network-wide state to make
coordinated security decisions. We give an example of each below.

\textit{P3: Distance-based access control:} This policy grants access to a service only if the user is within a certain distance
from a physical location (e.g., the server room); this requires performing arithmetic operations over GPS coordinates embedded in the packet header:

\begin{lrbox}{\mybox}%
\begin{lstlisting}
  if ((lat-x)*(lat-x)+(lon-y)*(lon-y) < D)
  then  fwd(server)
  else  drop
\end{lstlisting}
\end{lrbox}
\scalebox{0.8}{\usebox{\mybox}}
\vspace{-3mm}

\textit{P4: Allow access only if admin is online:} \sys can support coordinated, network-wide policies by monitoring security events of interest and
making decisions based on the result.
For instance, a policy might grant access to a service only if the admin is online:

\begin{lrbox}{\mybox}%
\begin{lstlisting}
  def adminlst = ["Bob", "Alice"]
  c = count(match(usr in adminlst))
  if match(c>0) then fwd(server)
\end{lstlisting}
\end{lrbox}
\scalebox{0.8}{\usebox{\mybox}}
\vspace{-3mm}

\noindent\textbf{Advanced policies.} Inspired by the literature of
``continuous authentication''~\cite{eberz-2017-continuous,wang-2017-ca,muhammad-2018-ca,alzubaidi-2016-behavioral},
we propose a set of advanced policies that use device context to detect subtle but
important indicators of potential attacks. Due to space constraints, we only describe the high-level policies, but not the programs.
\textit{P5: Block requests without explicit user interaction}, which denies access to
a sensitive service if all apps are running in the background and there is no user interaction with the touchscreen to trigger the request;
such requests are likely generated by malware. \textit{P6: Scrub traffic if UIs are overlapping}, which
forwards traffic through a middlebox if the context information shows that app UIs are overlapping---a potential sign
for UI hijacking~\cite{Fratantonio-2017-SP}. \textit{P7: Conduct deep packet inspection if camera/recorder is on}, which detects if sensitive
information is being leaked through an active camera/recorder app\cite{Aditya-2016-mobisys}.

\vspace{-2mm}
\subsection{Compilation}
\label{ssec:compilation}
\vspace{-1mm}

Next, we discuss how the \sys compiler processes the key language constructs and generates P4 implementations.

\begin{figure*}[t]
\centering\includegraphics[width=14cm]{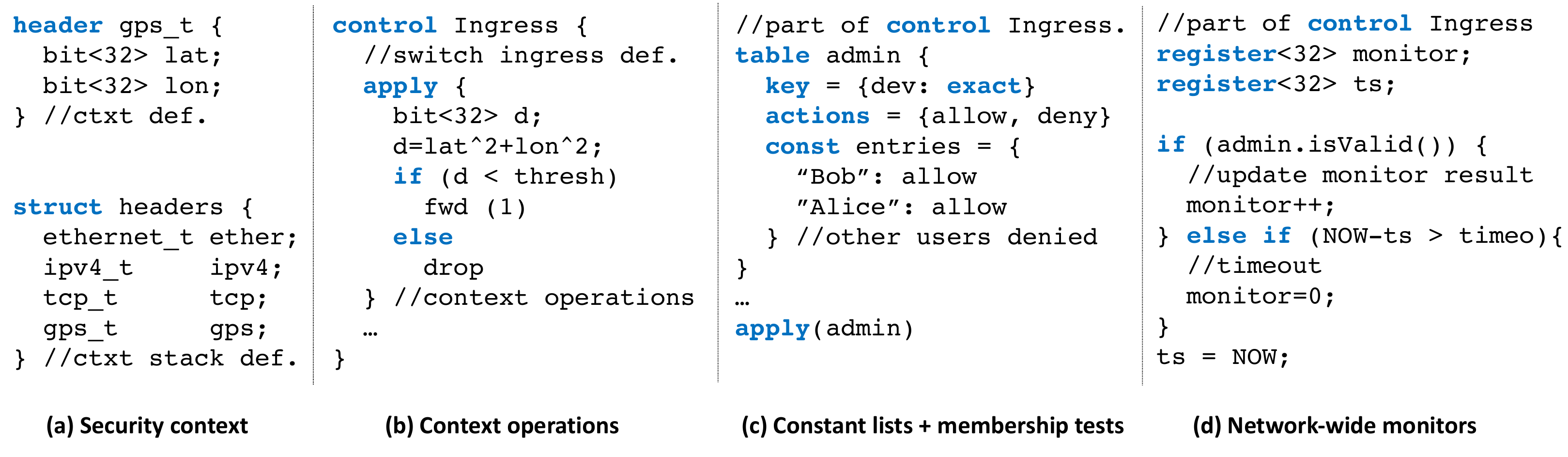}
\vspace{-1.5ex}
\caption{The \sys compiler processes the key language constructs and generates P4 implementations. The P4 snippets shown
are simplified for clarity of presentation. For instance, in (b), the instantiation of the \texttt{thresh} register is not shown;
in (d), the timestamp of a packet is obtained via the \texttt{ingress\_global\_timestamp} field
instead of a variable called \texttt{NOW}.}
\vspace{-3ex}
\label{fig:compilation}
\end{figure*}

\vspace{1mm}\noindent\textbf{Compiling security context.}
The \sys compiler generates custom header definitions for each
security context in a policy program. If a policy uses multiple types of context, the compiler generates
a stack of headers defined in P4 in the order they are used.
Context packets have special transport-layer protocol numbers, and context headers appear
right after transport-layer headers.
\sys does not modify normal traffic headers.
As a concrete example, Figure~\ref{fig:compilation}(a) shows the generated P4 snippets for the
\texttt{gps} header with two fields: latitude and longitude.

\vspace{1mm}\noindent\textbf{Compiling context operations.} The \sys compiler distinguishes between five classes of context operations:
arithmetic operations, bitwise operations, comparisons, context matches, and membership tests. The first three classes are simpler to handle, as
they can be directly translated into their P4 counterparts; the latter two require the compiler to generate additional code components in P4.
First off, all context fields are compiled into header definitions and references to these headers, as discussed above.
Then, for arithmetic, bitwise, or comparison operations over header fields, such as \texttt{lat*lat}, \texttt{sensors\&0x01}, or
\texttt{time<10}, our compiler forms expressions using the corresponding P4 operations over the headers.
For arithmetic operations, the current P4 specification supports addition, subtraction, and multiplication, which are all supported by
the \sys compiler. Notably missing from the list are division and modulo operations, which tend to be expensive to implement in switch hardware,
although sometimes they can be approximated by bit shifts if the divisor is a power of two.
If a \sys program involves operations unimplementable in P4, our compiler would reject the policy during compilation.

As an example, Figure~\ref{fig:compilation}(b) shows the P4 snippets that our compiler generates for computing the distance between a pair
of GPS coordinates to a pre-defined center (assumed to be \texttt{(0,0)}). Our compiler also generates
conditional statements based on the policy, e.g., \texttt{if-else} branches to test if the distance exceeds a threshold.
Context operations are performed within an \texttt{apply} block at the
\texttt{control Ingress}, which means the switch ingress pipeline.

Context matches, on the other hand, are compiled into match/action tables in P4. A match can be an \texttt{exact} match, which requires
matching a context field against a list of keys bit by bit. It could be a \texttt{lpm} match, which hits on the key that shares the longest
prefix. A match could also be \texttt{ternary}, which allows wildcards in the match keys. Ternary matches are performed
on TCAM (Ternary Content Addressable Memory), which is scarce switch resource.
\sys policy programmers do not have to specify which types of matches to use; rather, the compiler picks the best implementation for the policy.
Typically, context matches are performed against a user-specified constant list that defines membership, e.g., a set of devices owned by the
sales department. For a list with $k$ items \texttt{[a$_1$, a$_2$,$\cdots$, a$_k$]}, our compiler will construct a match/action table
with $k$ entries, where each entry corresponds to an item in the list. The actions associated with the entries depend on the mode of access
defined in the policy program.

For instance, consider the P4 snippet in Figure~\ref{fig:compilation}(c), which shows a match/action table generated according to a constant
list of two entries: Bob and Alice. The table implements an \texttt{exact} match on the device ID field. If the context match is successful, then
the device will be granted access to an enterprise resource; unsuccessful matches indicate that the context fails the membership test,
and needs to be denied access. Given that match/action tables consume most of the physical memory on chip, the compiler performs optimizations
for table layouts to minimize memory usage.

\vspace{1mm}\noindent\textbf{Compiling stateful monitors.} The \sys compiler generates a read/write register for each stateful monitor
in the policy, as well as code components for detecting monitored events and updating the monitor values. Such monitors are implemented as
a number of registers in P4, which are supported by SRAM on the switch. Updates to the registers are linespeed, so they can be performed
on a per-packet basis.
Specifically, for each incoming packet, the generated code checks whether this corresponds to an event of interest, using
either a context match, or a match over a membership list. If this event should be monitored, the code additionally updates
the monitor register and records the event timestamp. If a long time has elapsed after the previous event took place, then this register
is cleared to indicate that the monitored event is absent. As discussed before, monitors enable network-wide policies that make coordinated
security decisions---a policy can test if a monitor is active, and make decisions accordingly.

Concretely, the snippet in Figure~\ref{fig:compilation}(d) shows an example. It instantiates a 32-bit register to hold the monitor value, and
updates the register when the \texttt{admin} context is active in a packet. The code associates a timestamp to this monitor, and resets the
monitor upon timeout.

\vspace{1mm}\noindent\textbf{Compiling actions.}
An action is attached to each packet
to represent the final decision made on its context. In P4, decisions are represented by attaching special metadata fields to a
packet, which will be recognized and processed by a \textit{traffic manager}, which schedules packets to be sent on the correct outgoing
port(s) or dropped. Logging a packet is achieved by setting the outgoing port to be the switch CPU.

\vspace{1mm}\noindent\textbf{Optimizations.}
Programmable data planes have three types of notable constraints. \textit{Stages:} There is a fixed number of hardware stages,
and a packet can only match against one single context table per stage. \textit{Tables:} A single stage can only hold
a fixed number of tables. \textit{Memory:} Each stage has a limited amount of memory.

The \sys compiler performs three types of optimizations, which are particularly useful when composing multiple policies.
(a) If multiple policies check against the same context signal, our compiler will perform \textit{table deduplication}
to eliminate redundant context tables and save memory. (b) If policies use many small tables, \sys will perform \textit{table merge}
to reduce the number of needed tables. (c) If a policy performs more context checks than the number of available stages,
\sys will \textit{collapse} the policy  by \textit{recirculating} context packets to traverse the stages multiple times,
triggering different tables at each recirculation. This creates the illusion of a larger number of stages with
slightly increased latency.

\vspace{1mm} \noindent \textbf{Summary.} So far, we have described the basic compilation algorithm as if each packet is tagged with context information.
This makes it easy for a switch to access a packet's context without keeping state, but it results in high traffic overhead.
Next, we will relax this assumption by the design of a stateful, efficient, programmable in-network security primitive.

\vspace{-2mm}
\section{The In-Network Security Primitive}
\label{sec:primitive}
\vspace{-2mm}

At the core of \sys is a novel security primitive that is highly dynamic, efficient, and programmable.

\vspace{1mm}\noindent\textbf{Goal: A dynamic and efficient security primitive.}
The in-network primitive should ideally allow the level of protection to be adjusted between per-packet and per-flow granularities, by supporting
a tunable frequency of context packets for each connection.
At one end of the spectrum, per-flow granularity of protection degenerates into a static security mechanism that
does not support context changes within a connection. Thus the protection is very coarse-grained, especially for
long-lived connections that persist for an extended period of time (e.g., push-based mobile services, such as email~\cite{wang-2011-cellular}).
At the other end, per-packet granularity is extremely fine-grained, but may incur unnecessary resource waste unless
context changes from packet to packet. As a concrete example, if there are $20$ context fields across policies,
then each client needs to send $20\times 4/500=16\%$ extra traffic, assuming typical 500-byte packets and 4-byte context fields.
The \sys primitive supports a property that we call \textit{subflow-level security},
which achieves a tunable tradeoff between security granularity and overhead when enforcing context-aware security.

\vspace{1mm}\noindent\textbf{Property: Subflow-level security.}
We can state our desired property more formally below. Consider a sequence of packets in the same flow
$c_i, p_{i_1}, \cdots, p_{i_k}, c_{i+1}$, where $c$ represents a context packet and $p$ a data packet.
Subflow-level security requires that decisions made on the context packet $c_i$ should be applied to subsequent data
packets $p_{i_j}, i_j \in [i_1,i_k]$, but fresh decisions should be made for data packets that follow $c_{i+1}$.
The decision granularity can be tuned by $f$, the frequency of context packets. This results in an overhead
of $s\cdot f$, where $s$ is the size of context packets.
For instance, assuming $80$-byte context packets and a frequency of one context packet per ten seconds,
the overhead would be as low as $8$ bytes per second.

\vspace{1mm}\noindent\textbf{Challenges.}
Designing a primitive that supports subflow-level security, however, requires tackling three key challenges.
\textit{(a) Keeping per-flow state} requires a prohibitive amount of memory, but the SRAM of a modern switch is
on the order of $10$MB. Therefore, \sys approximates per-flow state using a on-chip key/value store.
\textit{(b) Buffering control plane updates} is necessary for handling new flows. Although context changes can be
entirely handled by the data plane, new flows require installing match/action entries from the switch CPU, which takes time.
Before updates are fully populated, \sys uses another hardware data structure akin to a cache to make conservative decisions
for buffered flows.
\textit{(c) Mitigating DoS attacks} that could arise due to the interaction between data and control planes. This defends against
malicious clients that craft special context packets to degrade the performance of selected clients or even the entire network.
In the next three subsections, we detail each of these techniques.

\vspace{-2mm}
\subsection{Approximating per-flow state}
\vspace{-1mm}

The key problem in the first challenge stems from the fact that the switch needs to process data packets without context attached to them.
Therefore, when a switch processes a context packet, it needs to remember the decision and apply it to subsequent data packets in the same
connection, until the next context packet refreshes the decision. A na\"{i}ve design would require keeping per-flow state on the switch.
This is infeasible because modern networks have large numbers of concurrent flows (about 10~million~\cite{miao-2017-silkroad}).
Keeping per-flow state requires about 140~MB switch SRAM (13-byte flow ID/five tuple, 1-byte decision for each flow), which
exceeds the memory capacity of the latest switch hardware (50-100~MB)~\cite{miao-2017-silkroad}.

\begin{figure}[t]
   \centering
   \includegraphics[width=0.3\textwidth]{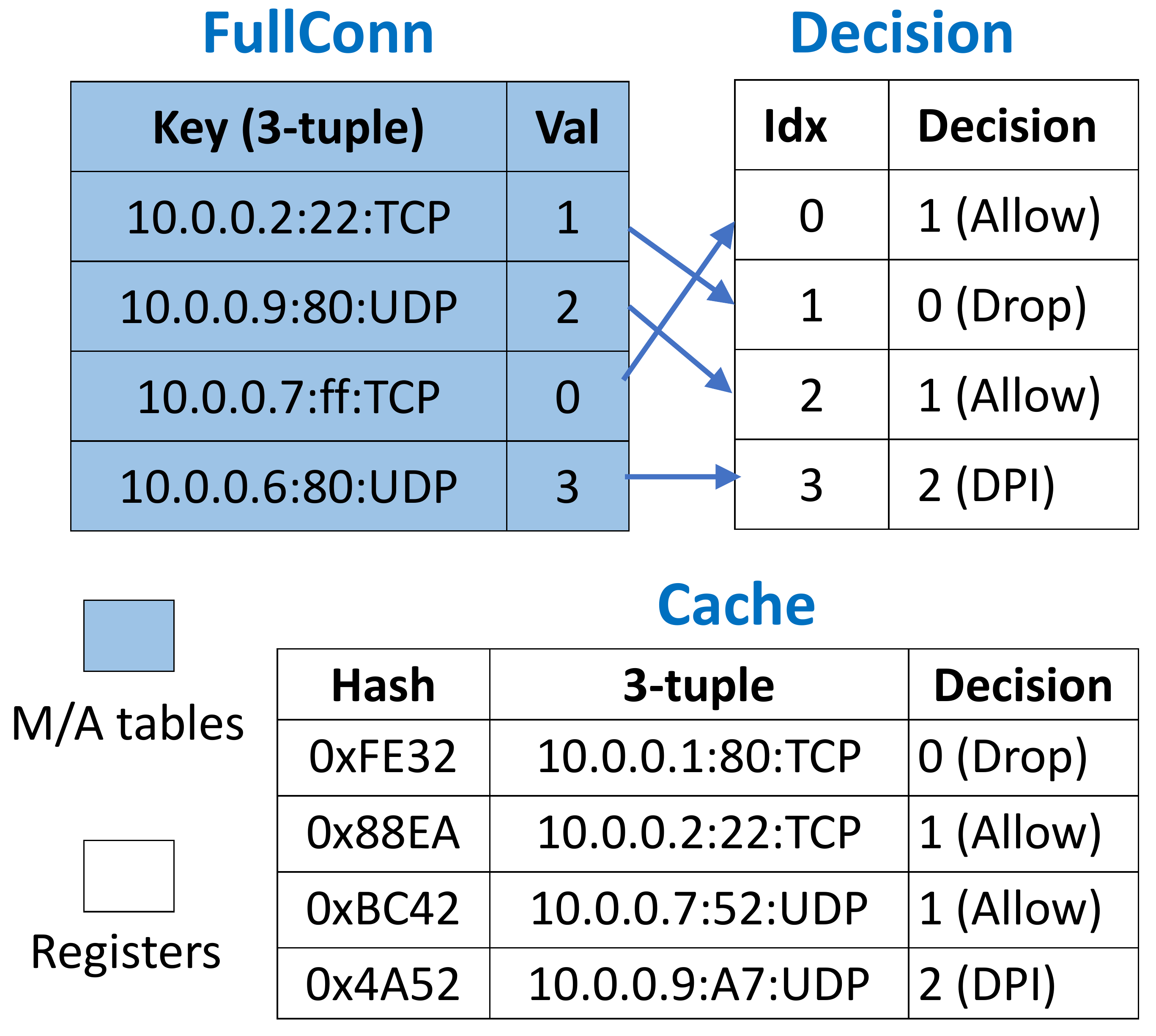}
   \vspace{-1ex}
   \caption{The key/value store with example entries.}
\vspace{-3ex}
   \label{fig:schema}
\end{figure}

To address this, \sys approximates per-flow state using a key/value store consisting of two data structures,
\texttt{FullConn} and \texttt{Decision}, as shown in Figure~\ref{fig:schema}.
The \texttt{FullConn} schema is \texttt{[sip, sport, proto]$\rightarrow$idx}.
The match key is the source IP/port and protocol for the client, and the value is an index
to a register array \texttt{R}. The indexed register \texttt{R[idx]} holds the decision made on the latest context packet within
this connection, and it can be refreshed entirely on the data plane. Insertions to this key/value store require control plane
involvement, but they are relatively infrequent and only needed for new connections.
Since the match key does not include the destination IP/port, this introduces some inaccuracy when a
client reuses a source port across connections. Therefore, for short-lived connections, data packets may see slightly outdated decisions.
To ensure that such inaccuracy does not misclassify a deny decision as an allow, we blacklist the source
IP addresses that have recently violated the enterprise policy: all connections from these clients would be blocked temporarily.

\vspace{-2mm}
\subsection{Buffering control plane updates}
\label{subsec:controlplane}
\vspace{-1mm}

Insertions to \texttt{FullConn} requires control plane involvement, so they take much longer than updating
policy decisions for an existing connection. As a result, when data packets in a new connection arrive at the switch,
the \texttt{FullConn} match/action table may not have been populated by the corresponding entry yet.
To address this, \sys uses a \textit{level of indirection} by creating
a small hardware \texttt{Cache} to buffer decisions for pending table updates,
which resides on the data plane and can be updated at linespeed.
All decisions in \texttt{Cache} are up-to-date, since writes to this cache are immediately effective, but this table has a smaller
capacity. The \texttt{FullConn} table takes more time to update, but it holds more connections.

\begin{figure}
\centering
\begin{subfigure}[b]{0.38\textwidth}
   \includegraphics[width=0.9\textwidth]{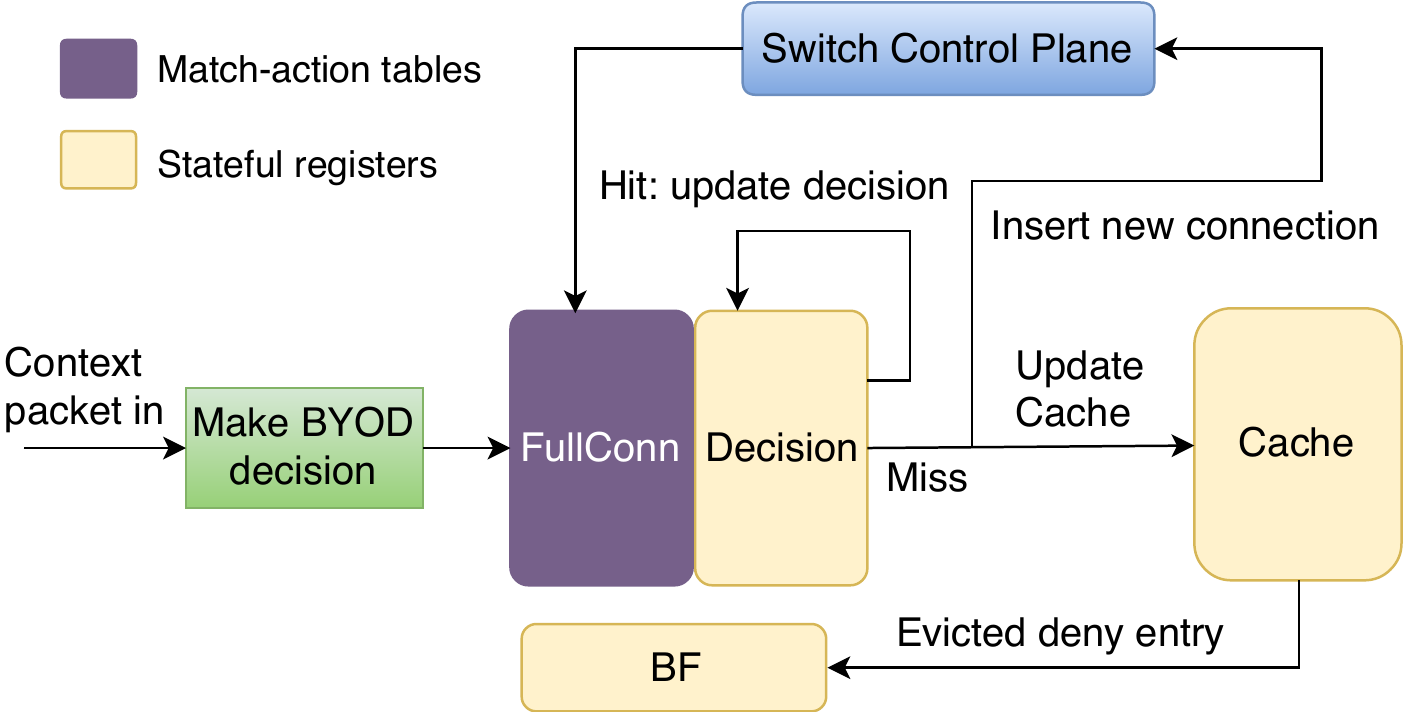}
   \vspace{-1ex}
   \caption{The logic for processing context packets}
   \label{fig:context}
\end{subfigure}
\begin{subfigure}[b]{0.38\textwidth}
	\vspace{1ex}
   \includegraphics[width=0.9\textwidth]{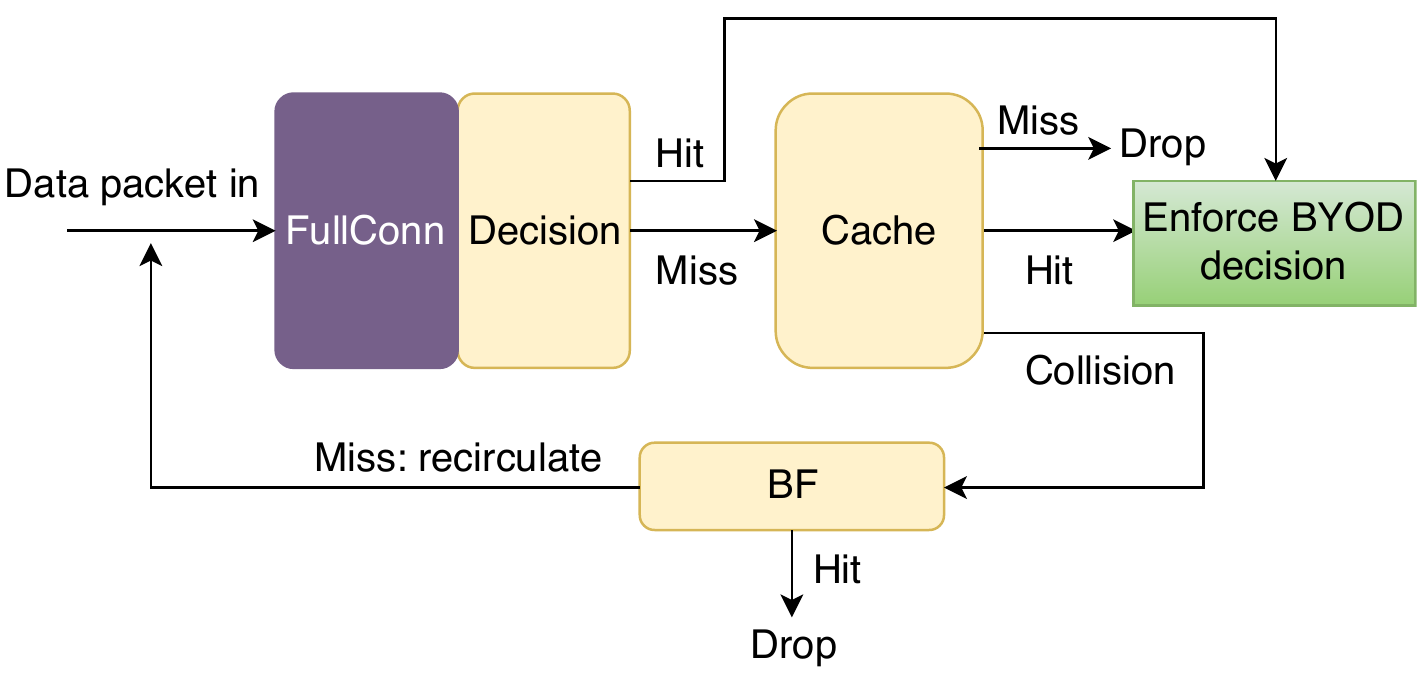}
   \vspace{-1ex}
   \caption{The logic for processing data packets}
   \label{fig:data}
\end{subfigure}
\caption{\sys uses a combination of match/action tables and stateful registers to process context and data packets.}
\vspace{-3ex}
\label{fig:dataflows}
\end{figure}

\vspace{1mm}\noindent\textbf{The cache design.}
As shown in Figure~\ref{fig:schema}, \texttt{Cache} has a fixed number of entries.
Our implementation uses $2^{16}$ entries, which corresponds to
the output size of a CRC-16 hash function. Each entry is of the form \texttt{h$\rightarrow$[sip,sport,proto,dec]}, where
\texttt{h} is the CRC hash of the flow's three tuple, i.e., \texttt{h}=CRC(\texttt{sip,sport,proto}), and \texttt{dec} is the
decision made based on the context packet. The size of \texttt{Cache} is $2^{16}\times (7+1)$=$0.38$~MB memory.
When \sys receives a context packet from a new connection (Figure~\ref{fig:context}), it immediately adds the entry to \texttt{Cache},
and then invokes the control plane API to insert the match/action entry in \texttt{FullConn}.
Since CRC functions are not collision resistant, different connections may be mapped to the same entry, and hence we evict old entries upon collision.
When a data packet comes in (Figure~\ref{fig:data}), \sys first matches it against the \texttt{FullConn} table and applies the decision upon success.
If there is no entry for this packet, then \sys indexes the \texttt{Cache} table instead.
Upon a cache hit, the corresponding decision is applied to the data packet. Upon a cache miss,
one of two situations has happened: a) the switch has not seen a context
packet from this client, or b) the entry for this client has been evicted due to the collision. \sys distinguishes between these cases
using the following cache eviction algorithm.

\vspace{1mm}\noindent\textbf{Handling cache evictions.}
Upon collision, we always replace the existing entry with the new one. This is because \sys has already invoked
the control plane to install the corresponding entry in \texttt{FullConn}, which will complete in time.
Therefore, if a packet does not match any entry in \texttt{FullConn} and experiences a collision in \texttt{Cache}, we use a special instruction to
recirculate the packet inside the data plane to delay its processing. Recirculated packets are sent back to the
switch ingress to be matched against the \texttt{FullConn} table one more time. This recirculation is repeated
up to $k$ times, where $k$ is chosen to be larger than the expected time for the control plane to populate an entry.
If a packet has reached this threshold, and the \texttt{FullConn} table still has not been populated. Then we consider this to
be case a) above, and drop the packet.

\vspace{1mm}\noindent\textbf{Early denies.}
To reduce the amount of recirculated packets, we make early decisions to drop a packet if its context is evaluated to a ``deny''.
Specifically, when evicting an entry from \texttt{Cache}, we add its source IP address into a blacklist Bloom filter
(\texttt{BF} in Figure~\ref{fig:dataflows}) if the decision is to drop.
Source addresses in \texttt{BF} represent devices that have violated the policy recently
and need to be blacklisted for a period of time. If a packet cannot find an entry in either \texttt{Cache} or \texttt{FullConn}, but
hits \texttt{BF}, we drop it without recirculation.
Since Bloom filters can only produce false positives, but never false negatives, we will always correctly reject an illegal connection.
However, we might err on the conservative side and reject legal connections as well, if the \texttt{BF} produces a false positive.
This is a rare case, however, as this will only happen during the window in which \texttt{FullConn} has not been populated,
the \texttt{Cache} entry has been evicted, and the \texttt{BF} happens to produce a false positive.
Nevertheless, \sys periodically clears this Bloom filter to reduce false positive rates, which grow with the number of contained elements.
When the \texttt{BF} is being cleared, packets will be recirculated until the operation completes.

\vspace{-2mm}
\subsection{Handling denial-of-service attacks}
\label{subsec:dos}
\vspace{-1mm}

Since \sys requires extra processing inside the network, we need to ensure that it does not introduce new attack vectors.
Specifically, we have identified two potential denial-of-service attack vectors and hardened the primitive against them.

\vspace{1mm}\noindent\textbf{Total residency attacks.} Different from stateless, IP-based routing,
\sys keeps state in the \texttt{FullConn} table. Therefore, an attacker could initiate many new connections
and try to a) overwhelm the \texttt{FullConn} table and b) constantly involve the switch CPU to install new entries.
Although the enterprise network can easily prevent spoofing, such attacks are still feasible by varying the source port and the protocol ID.
Therefore, \sys maintains the number of active connections per enterprise IP, and controls the growth of the \texttt{FullConn} table.
In addition, the \sys control plane periodically scans through the \texttt{FullConn} table and expires inactive entries (using hardware support)
to make room for new connections.

\vspace{1mm}\noindent\textbf{Cache eviction attacks.}
The above algorithm defends against a malicious attacker that generates many connections
to overwhelm the \texttt{FullConn} table. However, an attacker can also launch targeted DoS attacks without initiating a suspiciously
large number of connections. Specifically, she could send context packets more frequently than usual, and try to evict cache entries
from \texttt{Cache} that are mapped to the same bucket. Although the attacker may not know the hash seed, therefore cannot predict who would
be the victim of the attack, she could degrade the performance of the connection that shares the same hash entry, if one exists.
To prevent such attacks, we enhance the cache eviction strategy.
When replacing an old entry $e_o$ with a new entry $e_n$, we check whether these two entries are from the same source IP.
If so, we immediately replace the entries. If not, we opportunistically perform the replacement. By doing so, we limit the amount
of damage an attack can cause by sending frequent context packets.

\vspace{-2mm}
\section{Orchestrating \sys}
\label{sec:orchestration}
\vspace{-2mm}

Next, we explain how we orchestrate the \sys in-network primitive using a centralized \textit{controller},
and describe the \textit{client module} that runs on the mobile devices for context collection.

\vspace{1mm}\noindent\textbf{The \sys controller.}
\sys has a centralized controller that hosts the compiler and distributes the generated data plane programs to
the switches.
Unlike an SDN controller, which actively makes decisions on behalf of the data plane, the \sys controller is
not involved in packet processing, so it \textit{does not pose any software bottleneck}. The main controller runs in a remote server, and uses
well-defined RPC calls to communicate with programmable switches' local control planes. Each switch has a local control plane
that runs on the switch CPUs, and it configures the switch data plane
by installing match/action table entries, loads new switch programs, and serves as the primary logging component of the system.

\begin{figure}[t]
	\centering
	\includegraphics[width=0.45\textwidth]{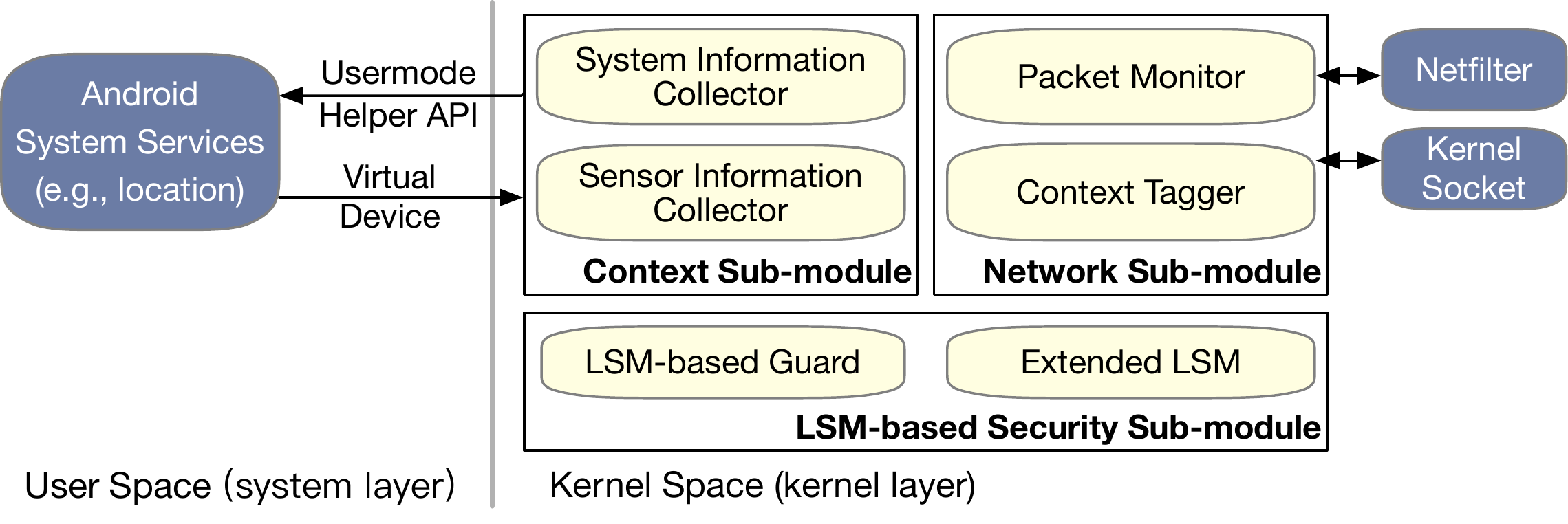}
\vspace{-2ex}
	\caption{The architecture of the PoiseDroid client module.}
\vspace{-3ex}
	\label{fig:context-collector}
\end{figure}

\vspace{1mm}\noindent\textbf{The \sys client module.}
Our client module PoiseDroid is installed at BYOD devices to collect context information and embed it into packets.
PoiseDroid does not require modification of existing Android apps, but rather acts as
a pre-positioned kernel module. Since most Android smartphones disallow users to load other kernel modules, PoiseDroid
cannot be removed or modified by mobile malware.
When the device connects to the enterprise network, our module uses TOTP~\cite{m2011rfc} for authentication,
although we note that \sys can be integrated with other types of authentication.
Figure~\ref{fig:context-collector} shows the architecture of PoiseDroid with three submodules.

\vspace{1mm}
\noindent\textit{The context submodule.}
It collects context information from the Android system services~\cite{Ye2017} using usermode-helper APIs~\cite{usemodeapi,dumpsys},
and it registers a virtual device to redirect the context data to our kernel module.
The information to be collected is specified by a generic BYOD client configuration, which includes
a) app information, such as UIDs of active apps,
b) system information, such as screen light status, and c) device status, such as accelerometer and gyroscope readings.
Note that the network may only use a subset of such context information for decision making.

\noindent\textit{The protection submodule.}
It protects the registered virtual device, the system tools (e.g., \texttt{dumpsys}), and the system services
using LSM hooks in Android kernel~\cite{backes2014android, nadkarni2014asm}.
It monitors invocations of selected system calls, such as \emph{ptrace()}, \emph{open()}, \emph{mprotect()} and \emph{chown()},
and prevents any other processes to write false data to these protected components.

\noindent\textit{The network submodule.}
It crafts and sends special context packets with signals needed for the enterprise policies, using a frequency specified in
the configuration. When an app opens a new socket, or when an existing socket sends packets after being dormant for a while,
it also generates a context packet.

\vspace{-2mm}
\section{Evaluation}
\label{sec:eval}
\vspace{-2mm}

In this section, we describe the experimental results obtained using our \sys prototype. Our experiments are designed to
answer four research questions:
a) How well does the \sys compiler work?
b) How efficiently can \sys process the security contexts inside the network?
c) How well does \sys scale to complex policies?
d) How much overhead does the \sys client incur on mobile devices?
and e) How does \sys compare with traditional SDN-based security?

\vspace{-2mm}
\subsection{Prototype implementation}
\vspace{-1mm}

We have implemented the \sys prototype using $4442$ lines of code in C/C++ and Python, and will release
the source code to the community.
The \sys \textit{compiler} is implemented in C++, using Bison 2.3 as the syntax parser, and Flex 2.5.35 as the lexer.
It can generate switch programs in P4 for the Tofino hardware.
The PoiseDroid \textit{client module} is implemented in C as a pre-positioned kernel module on Linux 3.18.31.
It extends the default LSM framework, SEAndroid, to implement the protection submodule.
For the evaluation, PoiseDroid runs on a Pixel smartphone with a Qualcomm Snapdragon 821 MSM8996 Pro CPU (4 cores) and Android v7.1.2.
The \sys \textit{control plane} is implemented in Python, and runs as part of the control plane software suite for the Tofino switch.
It manages the match/action table entries and reconfigures the data plane programs.
It can also be configured to invoke the hardware-based packet generator on the Tofino chip to send traffic
at full linespeed, which we have used to test the latency and throughput of \sys.

\vspace{-2mm}
\subsection{Experimental setup}
\vspace{-1mm}

We conducted our experiments on a hardware testbed with one Wedge 100BF Tofino switch and two servers.
The Tofino switch has a linespeed of 100~Gbps per port, and 32 ports overall, achieving an aggregate throughput of 3.2~Tbps when
all ports are active. 
Each server is equipped with a six-core Intel Xeon E5-2643 CPU, 16~GB RAM, 1~TB hard disk, and
four 25~Gbps Ethernet ports, which collectively can emulate eight forwarding decisions (one per server port).
The servers are connected to the Tofino switch using breakout cables from the 100~Gbps switch ports
to the 25~Gbps server Ethernet ports. At linespeed, the testbed should achieve full 100~Gbps bandwidth per switch port.

On the first server, one of its ports is configured to be an enterprise server, and other ports are configured to emulate a DPI device,
a traffic scrubber, and a logger, respectively.
The other server functions as an enterprise client.
The mobile traces are first collected from our Pixel smartphone, and then ``stretched'' to higher speeds to be replayed.
The replay can be initiated from
a) the enterprise client, or b) the hardware generator for \sys at linespeed.
We have performed functionality tests on both environments, although most of the reported results are obtained from the hardware testbed.

\vspace{-2mm}
\subsection{Compiler}
\label{subsec:compilation_speed}
\vspace{-1mm}

We start by evaluating the performance of the \sys compiler and its generated programs.

\noindent\textbf{Compilation speed.}
In order to understand the performance of our compiler, we measured the time it took to generate
switch programs for each of the seven policies. We found that compilation finished within one
second across all policies. P1 and P3 took slightly more time than the rest, because they involve more
context fields and our compiler needs to generate more logic for header processing.

\noindent\textbf{Generated P4 programs.}
The generated P4 programs for the policies have 855-975 lines of code, which are significantly more complex than the original policy programs that only contain a few lines of code.
The programs can support a maximum of 1 million connections. 
(SilkRoad~\cite{miao-2017-silkroad} reports a maximum of 10M connections in datacenters and switch memory
ten times larger than our low-end Tofino switch; we have therefore scaled down the number of connections proportionally.)
The most important takeaways are a) utilization for all types of resources is low, and b) policy P4 requires relatively
more ALUs due to the monitor update logic.
The utilization for SRAM (used for exact match) is roughly 43\%, for TCAM (used for longest-prefix match)
is below 1.1\%, and for VLIWs (Very Long Instruction Words, used for header modifications) is below 7\%.
In other words, \sys leaves plenty of resources on the switch to support other functionalities.

\vspace{-2mm}
\subsection{In-network processing overhead}
\label{subsec:in-network-overhead}
\vspace{-1mm}

Next, we turn to evaluate the overhead of \sys in terms of packet processing latency and switch throughput.

\begin{figure}[t]
\centering\includegraphics[height=3.5cm]{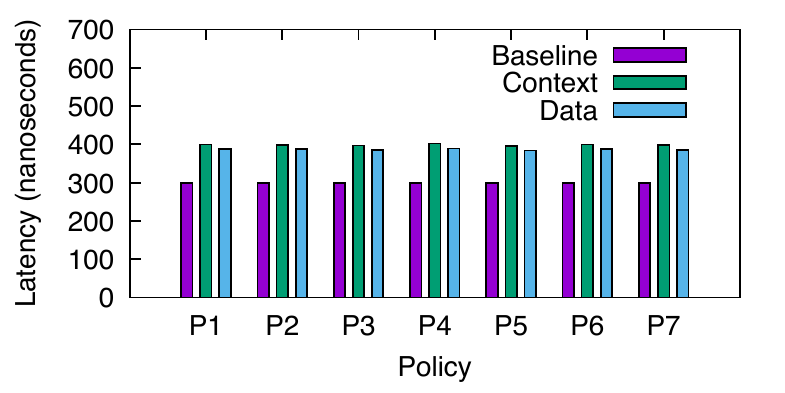}
\vspace{-2ex}
\caption{The amount of processing latency of \sys is small.}
\vspace{-1ex}
\label{fig:delay}
\end{figure}

\noindent\textbf{Packet processing latency.}
\sys adds to the overhead of packet processing, since it needs to check packet context and update
its tables to maintain per-flow state. To quantify this overhead, we have tested the latency for \sys to
process a) a context packet, b) a data packet, and compared them with c) the latency for directly forwarding a non-\sys
packet without any processing. Figure~\ref{fig:delay} shows that for all tested policies,
the additional latency is 88 nanoseconds for processing data packets, and 100.4 nanoseconds for
processing context packets. In an enterprise network where the round-trip delays are on the order of milliseconds,
such a small extra latency is negligible.

\begin{figure}[t]
\centering\includegraphics[height=3.5cm]{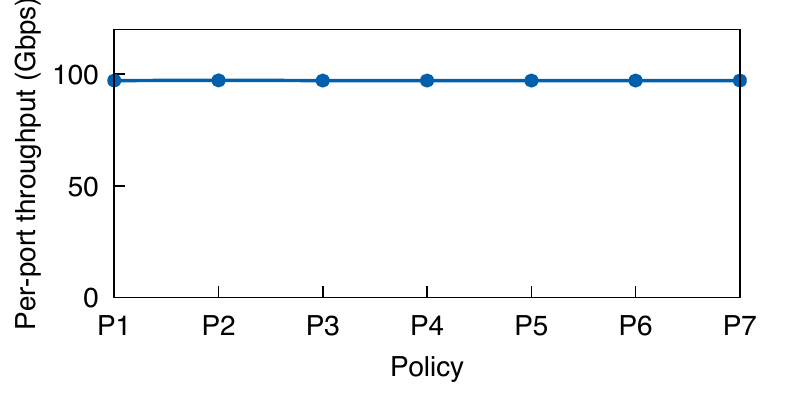}
\vspace{-2ex}
\caption{\sys achieves full linespeed programmability.}
\vspace{-2ex}
\label{fig:throughput}
\end{figure}

\noindent\textbf{Switch throughput.}
Next, we measured the throughput per switch port. Since the linespeed of the port is very high, we used
the hardware packet generator instead of the servers for stress testing. The generator ingested mobile traces collected
from our phone, and stretched the trace to be 100~Gbps. Figure~\ref{fig:throughput} shows the per-port throughput
for all policies. As we can see, although there is additional processing delay in \sys, the pipelined nature of the
switch hardware makes it achieve full bandwidth nevertheless. In other words, \sys leverages programmable data
planes to achieve linespeed programmability, a key goal that we have designed for.

\vspace{-2mm}
\subsection{Scalability}
\vspace{-1mm}
\label{sec:scalability}

\begin{figure*}[t]
\centering
\subcaptionbox{Num. of contexts vs. num. of range checks\label{fig:ctx-num-entry}}
      [.33\linewidth]
      {\includegraphics[height=3.5cm]{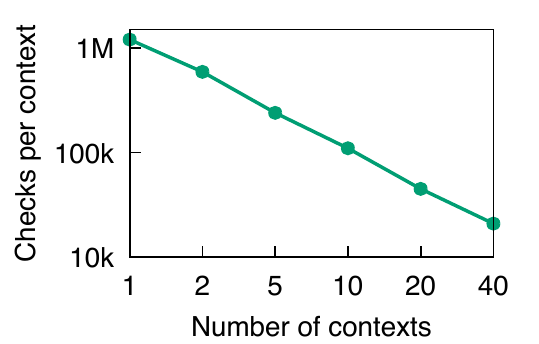}}
\subcaptionbox{Num. of contexts vs. latency\label{fig:ctx-num-delay}}
      [.33\linewidth]
      {\includegraphics[height=3.5cm]{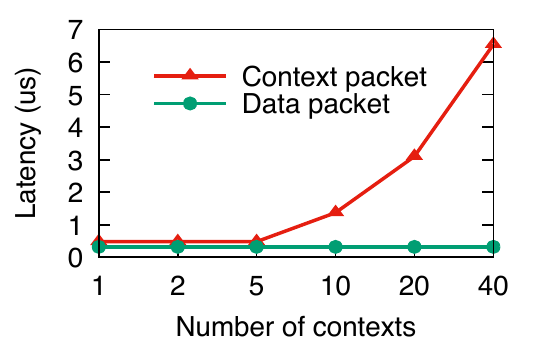}}
\subcaptionbox{Num. of contexts vs. traffic overhead\label{fig:ctx-num-throughput}}
      [.33\linewidth]
      {\includegraphics[height=3.5cm]{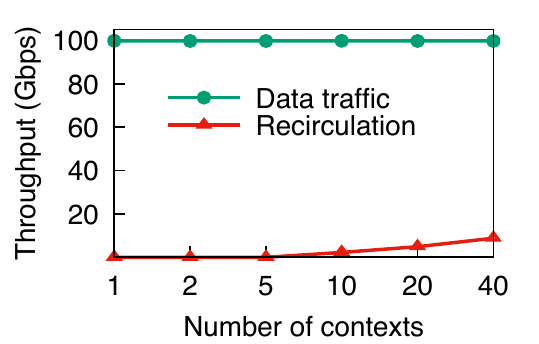}}
\vspace{-1ex}
\caption{\sys can perform 1.2 million concurrent range checks for a single context, or 21k concurrent range checks for a maximum of 40 context types.
Context packets with more than 5 context types need to be recirculated multiple times; \sys supports a maximum of 8 recirculations, which
leads to 6$\mu$s latency and $8.8$~Gbps additional traffic in a dedicated recirculation pipeline. Data packets are not affected by policy complexity,
as they simply look up the decisions from the connection table.}
\vspace{-2ex}
\label{fig:check-perf}
\end{figure*}

Next, we evaluate how \sys scales to complex policies. We note that the ``number of policies'' is an inexact metric, as policies may perform different
numbers of checks on different numbers of context types. Therefore, our methodology is to instantiate a large number of ``unit policies'',
each of which only performs a single check on a single context. We then use the compiler to compose them together, and analyze
the complexity of the composed policy in two dimensions: a) the number of checks per context, and b) the number of context types.
For instance, consider the following unit policies:

\begin{lrbox}{\mybox}%
\begin{lstlisting}
  if match (time>10) then fwd(mbox)
  if match (time>15) then drop
  if match (lib==1.0.2) then fwd(server)
\end{lstlisting}
\end{lrbox}
\scalebox{0.8}{\usebox{\mybox}}
\vspace{-3mm}

\noindent
We say that the composed policy has two context types and performs three checks---1.5 checks per context on average.
\noindent\textbf{Number of range checks.}
\sys compiles each range check into a match/action entry, so the number of checks a switch can suppprt
depends on its available memory. We first measured the maximum number of checks \sys can perform on a single context,
by asking the compiler to compose more and more unit policies until the compilation failed.
We found that our switch can support 1.2 million concurrent checks, which are spread across 5 hardware stages
(memory in other stages are taken by the 1M connections).
High-end Tofino switches have ten times as much memory~\cite{miao-2017-silkroad};
which would allow proportionally many concurrent checks.

\noindent\textbf{Number of context types.}
\sys compiles each context type into a match/action table, so the number of context types is bottlenecked
by the number of tables that a switch can support. We increased the variety of context types (e.g., time, library version)
from one to the maximum before compilation failure, and for each data point, we measured the maximum number of
checks \sys can perform per context. We found that each of the 5 stages can support 8 context tables,
so \sys supports a maximum of 40 context types. The number of checks per context decreases as we added more context types,
and the lowest number is 21k checks per context (Figure~\ref{fig:ctx-num-entry}).
High-end switches have 2--3 times as many hardware stages,
so they would be able to support 80--120 context types.

\noindent\textbf{Overhead.} We define a ``baseline'' to be the latency and throughput for a unit policy, where a context packet
traverses the hardware stages exactly once without recirculation. A packet with $k$ context types would recirculate
$\lceil \frac{k}{5} \rceil$ times, since every traversal would match 5 tables, one in each available stage.
At the maximum, \sys supports 8 recirculation for 40 context types, leading to a latency of 6$\mu$s (Figure~\ref{fig:ctx-num-delay}),
which is still orders of magnitude lower than typical enterprise RTTs (ms).
Recirculation also causes extra traffic overhead. As Figure~\ref{fig:ctx-num-throughput} shows,
the maximum recirculation traffic overhead is 8.8Gbps for the switch.
Since switches have dedicated pipelines to handle recirculation traffic, 
regular switch ports can still forward traffic at linespeed.

\noindent\textbf{Discussion.} One might also wonder how many devices \sys can support. However, the ``number of devices''
is also an inexact metric, as \sys supports per-device policies simply by including the device ID as a context type.
Therefore, supporting per-device policies merely reduces the number of total context types by one, from 40 to 39.
The number of unique checks \sys can support---or the number of device IDs---is 21k with 39 context types.
As another dimension of constraint, assuming each device may launch 1k concurrent connections, then \sys would support
a maximum of 1M/1k=1k devices.

\vspace{-2mm}
\subsection{Client overhead}
\label{subsec:clientoverhead}
\vspace{-1mm}

We now evaluate the overhead of the client module, using the vanilla Android without PoiseDroid
as the baseline system.

\noindent\textbf{CPU overhead.}
We tuned the frequency at which the client module sends
context packets, and measured the CPU overhead for each frequency.
In each trial, we uploaded a video file
of 1.73~GB to a remote FTP server using the mobile app \texttt{AndFTP}~\cite{andftp},
and measured the CPU overhead as collected from the  \texttt{/proc/loadavg} file.
In a na\"{i}ve design where PoiseDroid tags every packet
with context information, the CPU overhead is as much as 11\%. With an optimized design where the client module
tags one packet with context per second, the CPU overhead is drastically reduced to 1.3\%.

\noindent\textbf{Traffic overhead.}
Next, we tested the traffic overhead due to the context packets.
This experiment assumes four context fields (16~bytes).
We found that, at one context packet per second, the traffic overhead is
less than 0.01\%, a negligible amount.

\noindent\textbf{Battery overhead.}
We used PCMark~\cite{pcmark}, a battery life benchmark tool to test smartphones and tablets, to quantify the amount of battery overhead.
PCMark tests capture a wide variety of activities, such as browsing, video playback, photo editing, writing, and data manipulation.
In the beginning of the experiment, the phone was charged with full capacity (100\%), and the tests ran until the battery dropped to less than 20\%.
The overall overhead across activities introduced by PoiseDroid is only 1.02\%, and even for the
activities that introduce the highest overhead (i.e., writing), the overhead is only 2.87\%.

\noindent\textbf{Overall benchmark.}
Next, we used CF-Bench, a comprehensive benchmark tool designed for multicore mobile devices, to quantify the overall overheads of PoiseDroid.
This tool can further measure the overheads introduced by native code, Java code, and an overall benchmark score, where higher scores mean
better performance.
PoiseDroid only introduces 5\%, 4\%, and 5\% additional overhead for the native, Java, and overall scores.

\vspace{-2mm}
\subsection{\sys vs. OpenFlow-based SDN}
\vspace{-1mm}

Last but not least, we compare the paradigm of programmable in-network security, as embodied in \sys, against the paradigm of
OpenFlow-based SDN security, in terms of a) resilience to control plane saturation attacks~\cite{shin-2013-avantguard}, and b) agility in security posture change.
As we motivated in \S\ref{sec:overview}, one key advantage of \sys over traditional SDN security solutions
is the avoidance of software-based packet processing on a remote controller, because \sys uses programmable data planes
to directly process context information in the hardware fast path.

\vspace{1mm}
\noindent\textbf{Setup.}
We set up a Floodlight \texttt{v1.2} SDN controller on a separate server, and configured other servers to use
the controller using OpenFlow as implemented in OpenvSwitch \texttt{v2.9.2}.
We implemented our example policies (P1-P7) as software ``SDN apps'' in the controller.
These apps listen for client context updates, and push OpenFlow rules to the clients for access control.
This closely mirrors the setup in state-of-the-art
security solutions based on OpenFlow-based SDN~\cite{oconnor-2018-pivotwall,shin-2013-avantguard,hong-2016-pbs}.

\begin{figure}[t]
\subcaptionbox{End-to-end latency\label{fig:sdn-delay}}%
      [.4\linewidth]
      {\includegraphics[height=3.5cm]{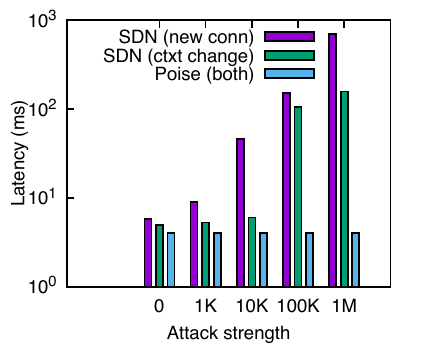}}
\hspace{1.8em}%
\subcaptionbox{Successful connections\label{fig:sdn-throughput}}%
      [.4\linewidth]
      {\includegraphics[height=3.5cm]{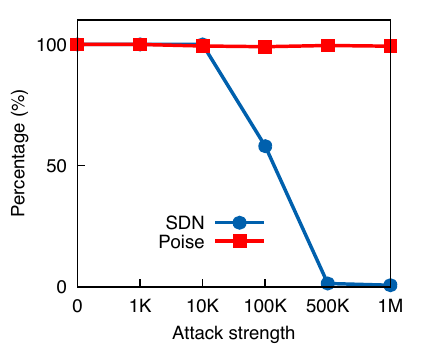}}
\caption{\sys is resilient to control plane saturation attacks. Attack strength is measured by the number
of new connections (or context changes for existing connections) per second. New connections trigger
\texttt{PacketIn} and \texttt{FlowMod} events, and context changes trigger \texttt{FlowMod} events.}
\vspace{-3ex}
\label{fig:sdn}
\end{figure}

\vspace{1mm}
\noindent\textbf{Control plane saturation attacks.}
We launched attacks with different strengths, as measured by the number of new connections per second~\cite{shin-2013-avantguard}.
As Figure~\ref{fig:sdn} shows, since each new connection triggers a  \texttt{PacketIn} event at the controller,
the central controller was completely inundated
with an attack strength of 1M connections per second. During the attack, concurrent legitimate clients were not able to establish
new connections (99\%+ connection requests from legitimate clients were dropped;
the rest experienced a latency $122\times$ higher than normal).
\sys, however, paints a completely different picture---although new connections will trigger the installation of match/action entries
from the switch CPU, they are simultaneously added to the hardware cache at linespeed, so
its performance stays \textit{almost constant} during the attacks.

\vspace{1mm}
\noindent\textbf{Defense agility attacks.} We quantify the \textit{defense agility} of a security system by
measuring $\delta$, the time it takes to change its security posture (e.g., access control decision)
after seeing a new context packet. In OpenFlow-based SDN, this incurs a round-trip time delay
for the context packet to reach the controller and for the controller to push new OpenFlow rules via \texttt{FlowMod} messages.
We found that, depending on the network load, the agility of the baseline system is $\delta=$5~ms--2.47~s.
In comparison, \sys directly processes context changes on the fast path, achieving $\delta<500$~ns in all cases,
which is \textit{three to seven orders of magnitude} more agile.

This lag in adjusting security posture results in a \textit{window of vulnerability} that can be easily exploited by an attacker.
To demonstrate this, we have launched an attack as follows: a) the attacker launches a control plane attack in order to inflate $\delta$;
b) it then goes through a context change and sends an illegal request to the server; this request would eventually be blocked
when the OpenFlow rules are populated from the controller to the client, but this decision change is delayed due to the attack;
and c) as a result, the insecure request is incorrectly accepted by the enterprise server and access is granted to this attacker.
We then repeated the same attack to \sys, and found that it successfully blocked all such requests.

\vspace{1mm}\noindent\textbf{Summary.} These experiments demonstrate that \sys's ability to change security posture at linespeed
brings about significant security benefits. Compared to traditional SDN security solutions, the paradigm of programmable in-network security
not only increases the resilience to control plane saturation attacks, but also drastically boosts the defense agility.

\vspace{-3.5mm}
\section{Related Work}
\label{sec:related}

\noindent\textbf{SDN/NFV security.}
SDN/NFV-based solutions for enterprise security started with
SANE~\cite{casado-2006-sane} and Ethane~\cite{casado-2007-ethane}.
Recent followup work also include
PSI~\cite{yu-2017-psi} that uses virtualized middleboxes for security,
FortNox~\cite{porass-2012-fortnox} that supports role-based authorization,
Bohatei~\cite{fayaz-2015-bohatei} that defends against DDoS attacks,
PBS~\cite{hong-2016-pbs} that uses an SDN-like design for security enforcement,
PivotWall~\cite{oconnor-2018-pivotwall} and OFX~\cite{sonchack-2016-ofx} that perform information flow control using software controllers,
and CloudWatcher~\cite{shin-2012-cloudwatcher} that provides security monitoring as a service.
Existing work has also considered new attack vectors in SDNs~\cite{skowyra-2018-topotampering,
hong-2015-topoguard,shin-2013-avantguard,xu-2017-conguard}, such as control plane saturation attacks~\cite{shin-2013-avantguard}.
In comparison, \sys leverages programmable data planes for linespeed security,
increasing the defense agility and resilience to control plane saturation attacks.

\vspace{1mm}
\noindent\textbf{Context-aware security.} Security researchers have recognized the need for context-aware security to support
fine-grained, dynamic policies. Barth et al.~\cite{barth-2006-contextualintegrity} propose a logic framework for contextual integrity.
Recent work has developed various applications leveraging this concept.
ContexIoT~\cite{jia-2016-contextiot} analyzes UI activities, app information, and control/data flow information,
and prompts users for runtime permissions. FlowFence~\cite{fernandes-2016-flowfence} runs applications in sandboxes and enforces information flow control
across IoT applications. PBS~\cite{hong-2016-pbs} uses SDN software switches for BYOD security.
Yu et al.~\cite{yu-2016-iotsec} sketch a vision for using network function virtualization for context-aware IoT security.
DeepDroid~\cite{wang-2015-deepdroid} traces IPC and system calls to achieve fine-grained security.
Compared to existing work, \sys designs a network primitive for security enforcement,
and has an end-to-end framework for specifying, compiling, and enforcing declarative policies.

\vspace{1mm}
\noindent\textbf{Policy languages.}
Most domain-specific languages for networking~\cite{anderson-2014-netkat,
schelsinger-2014-concurrentnetcore,yu-2017-psi,voellmy-2010-nettle,monsanto-2013-pyretic,reitblatt-2013-fattire,
yuan-2015-netegg,beckett-2016-propane} are not targeted at security.
Policy languages for network security also exist, but we are not aware of an existing language that can support
context-aware policies on programmable data planes. For instance,
PSI~\cite{yu-2017-psi} uses finite state machines to specify security policies, but it assumes that the policies are implemented
by general-purpose virtual machines;
PBS~\cite{hong-2016-pbs} assumes a traditional SDN environment.
\sys builds upon an existing SDN language (NetCore~\cite{monsanto-2013-pyretic}), but adapts it for
enforcing context-aware security on programmable data planes.

\vspace{1mm}
\noindent\textbf{Programmable data planes.} \sys builds upon the emerging trend of using data plane
programmability~\cite{bosshart-2013-metamorphosis,bosshart-2014-p4,song-2013-pof} for in-network computation,
e.g., load balancing~\cite{katta-2016-hula}, network monitoring~\cite{narayana-2017-marple},
key-value cache~\cite{jin-2017-netcache,liu-2017-incbrick}, and coordination~\cite{jin-2018-netchain},
but it focuses on a very different goal: security.
The closest to our work is a recent position paper~\cite{morrison-2018-poise},
but it neither has a full system implementation nor evaluation.

\vspace{-2.5mm}
\section{Conclusion}
\label{sec:conclusion}
\vspace{-2.5mm}

We have described \sys, a system that can enforce context-aware security using a programmable, efficient, in-network primitive.
In \sys, administrators can express a rich set of policies in a high-level language. Our compiler then
compiles the policies down to switch programs written in P4. These programs run inside modern switches with
programmable data planes, and can enforce security decisions at linespeed.
Our evaluation shows that \sys has reasonable overheads, and that compared to OpenFlow-based defense,
it is highly agile and resilient to control plane saturation attacks.


\bibliographystyle{abbrv}
\bibliography{paper}

\end{document}